\documentclass[twoside,twocolumn,9pt]{article}
\usepackage{extsizes}
\usepackage[left=1.5cm, right=1.5cm, top=1.785cm, bottom=2.0cm]{geometry}
\usepackage{balance}
\usepackage{times,mathptmx}
\usepackage{sectsty}
\usepackage{amsmath}
\usepackage{authblk}
\usepackage{graphicx}
\usepackage{abstract} 
\usepackage{lastpage}
\usepackage[format=plain,justification=justified,singlelinecheck=false,font={stretch=1.125,small,sf},labelfont=bf,labelsep=space]{caption}
\usepackage{float}
\usepackage{fancyhdr}
\usepackage{fnpos}
\usepackage[english]{babel}
\addto{\captionsenglish}{%
  
}
\usepackage{array}
\usepackage{charter}
\usepackage{droidsans}
\usepackage[T1]{fontenc}
\usepackage[usenames,dvipsnames]{xcolor}
\usepackage{setspace}
\usepackage[compact]{titlesec}
\usepackage{hyperref}

\definecolor{cream}{RGB}{222,217,201}

\newcommand{\kB}{k_\mathrm{B}}

\newcommand{\Ea}{E_\mathrm{b}}

\title{Morphology Formation in Binary Mixtures upon Gradual Destabilisation}
\author[1]{Charley Schaefer\footnote{Corresponding author: Charley.schaefer@york.ac.uk}}
\author[2]{Stefan Paquay}
\author[1]{Tom C. B. McLeish}
\affil[1]{Department of Physics, University of York, Heslington, York, YO10 5DD, UK}
\affil[2]{Martin Fisher School of Physics, Brandeis University, Waltham, MA, USA}
\date{  }

\begin{document}
\maketitle

\begin{abstract}
Spontaneous liquid-liquid phase separation is commonly understood in terms of phenomenological mean-field theories.
  These theories correctly predict the structural features of the fluid at sufficiently long time scales and wavelengths.
  However, these conditions are not met in various examples in biology and materials science where the mixture is  slowly destabilised, and phase separation takes place close to the critical point.
  Using kinetic Monte Carlo and molecular dynamics simulations of a binary surface fluid under these conditions, we show that the characteristic length scale of the emerging structure decreases, in $2$D, with the $4/15$ dynamic critical exponent of the quench rate rather than the mean-field $1/6$th power.
  Hence, the dynamics of cluster formation governed by thermodynamically undriven Brownian motion is much more sensitive on the rate of destabilisation than expected from mean-field theory.
  We discuss the expected implications of this finding to $3$D systems with ordering liquid crystals, as well as phase-separating passive or active particles.
\end{abstract}

\section{Introduction}

In fields of research ranging from biophysics \cite{ShinY17} to materials science \cite{Brabec10}, the theoretical modelling of phase-separating mixtures aims to relate the typical time and length scales of a phase-separated morphology to the physical properties of the molecular constituents as well as to the processing conditions \cite{deGrootBook84, Hansenbook06, Binder87}. 
In the classical picture, a homogeneous mixture of unlike molecules is initially stable, but upon a sudden temperature quench spontaneously demixes to form a phase-separated morphology that subsequently coarsens and arrests.
Phenomenological phase-field models are highly successful in explaining how coarsening and arresting are affected by viscoelastic effects \cite{Tanaka00}, hydrodynamics\cite{Furtado06, Krishnan15}, chemical reactions \cite{Glotzer95, Furtado06, Singh12, Krishnan15}, turbulent flow \cite{Perlekar14}, and the presence of a surfactant \cite{Benzi11}.
However, there is limited understanding of how the early-stage structure development is affected by the physical properties of the molecules and by the processing conditions \cite{Huston66, Nauman88, Wodo12, Schaefer15, Schaefer16, Schaefer18, Kessler16,QiS18, Cirillo19}.
Indeed, in case that the mixture is gradually destabilised, e.g., in case of slow cooling or of an ongoing concentration increase due to solvent evaporation, mean-field theory fails\cite{Carmesin86} and is unable to predict the rather strong dependence of the coarseness of the morphology on the quench rate  \cite{Schaefer18}.
In the present work, we rationalise this observation in terms of universal critical dynamics  \cite{Hohenberg77}.

According to mean-field theory, a demixed structure spontaneously emerges if the spinodal branch in the unstable region of the phase diagram is crossed.
Indeed, under these conditions there is no activation barrier and heterogeneities are amplified as time proceeds.
The fastest growing wavelength is sufficiently short for fast diffusion and sufficiently large to overcome surface tension.
If the mixture is, from initially stable conditions, slowly and steadily quenched into the phase diagram, the emerging length scale decreases with the square root of time.
At the point in time where the largest amplitude grows faster than the quench rate, rapid phase separation takes place.
Regardless of whether mean-field theory is corrected for thermal fluctuations according to the fluctuation-dissipation theorem  \cite{Cook70, Langer71, Petschek83, Lutsko12, Schaefer16}, the emerging length scale is predicted to decrease with the one-sixth power of the quench rate \cite{Huston66, Nauman88, Schaefer15, Schaefer16, Kessler16}.

Recently, kinetic Monte Carlo (kMC) simulations of a $2$D lattice fluid seemed to indicate that liquid-liquid demixing upon a gradual concentration quench leads to a one-fourth power, rather than one sixth  \cite{Schaefer18}.
This power-law exponent was interpreted to be originated by the diffusion of material towards nuclei: 
Phase separation takes place at the point in time where diffusion becomes faster than the rate at which the mean-free path becomes shorter, due to an increasing concentration.
While this argument compares the time scale of diffusion-limited nucleation to the quench rate \cite{Schaefer18}, the power one fourth may also be obtained by comparing the time scale of reaction-limited nucleation  to the quench rate \cite{Buil00}.
In that case, the size of the clusters is not coupled to the mean-free path, but to the stability of spherical critical nuclei as described by classical nucleation theory \cite{KashchievBook00}.
Although both theories seem to agree with the observations, the diffusion-limited nucleation argument seems to be specific to the case studied \cite{Schaefer18}, and a simple classical nucleation theory is inaccurate near the critical point \cite{Binder80}. 

In the present work, we argue that the length scale of a mixture that is slowly destabilised is not determined after the miscibility gap is entered, but actually prior to that, namely by universal critical dynamics \cite{Hohenberg77}.
We will support this discussion by conducting kinetic Monte Carlo (kMC) and molecular dynamics (MD) simulations of a binary molecular fluid that demixes in two spatial dimensions upon a gradual concentration and temperature quench.
Indeed, for the temperature-quenched mixtures, we recover the expectation that in the single-phase region the correlation length increases with the dynamic critical time exponent $4/15$ \cite{Hohenberg77, Alexander94, Godreche04}.
Since the time available for coarsening is $\Gamma^{-1}$, with $\Gamma$ the quench rate, the correlation length depends when entering the miscibility gap is $\Gamma^{-4/15}$, close to the earlier reported power of one fourth.

In the following, we present the theory of demixing from the phenomenological (mean-field) point of view, and then present two methods to go beyond the mean-field approximation, namely using kinetic Monte Carlo simulations of lattice model and using Langevin dynamics of a lattice-free molecular model.
Subsequently, we present our simulation results and briefly discuss the implications of our findings to a broader range of applications.

\section{Theory}

\subsection{Mesoscopic mean-field modelling} \label{sec:mean-field}

The phase separation of a homogeneous mixture with a single phase into a heterogeneous mixture with coexisting phases is thermodynamically understood by a favourable change in free energy.
Phenomenogically, an order parameter $-1 \leq \psi \leq 1$ can be assigned to the system (which could be related to the local concentration, magnetisation, orientation vector, etc.), and a dimensionless Landau free-energy density may be assumed\footnote{The energies are given in units of thermal energy $\kB T$, with $\kB$ Boltzmann's constant and $T$ the absolute temperature. The spatial positions/distances are given in units of the smallest length scale in the system, which is in our case the size of a molecule.}
\begin{equation}
f(\psi) = -\frac{1}{2}a\psi^2 + \frac{1}{4}b\psi^4.
\end{equation}
Indeed, coexisting phases with $\psi = \pm \sqrt{b/a}$ have a combined free energy $[f(+\sqrt{b/a})+f(-\sqrt{b/a})]/2$ smaller than that of the homogeneous mixture $f(\psi=0)$.
Here, $a>0$ and $b>0$ are phenomenological parameters, to which a physical meaning can be attributed by choosing an appropriate mean-field theory.

While phase separation is driven by the local free energy, it is counteracted by a free-energy penalty for the formation of spatial inhomogeneities of the order parameter (i.e., surface tension).
For sufficiently weak spatial gradients, this can be captured using a Van der Waals-type non-local free energy $\kappa/2 |\nabla \psi(\mathbf{r})|^2$ \cite{Hansenbook06}.
Integration over the system volume gives the Ginzburg-Landau free-energy functional 
\begin{equation}
  \mathcal{F} = \int \mathrm{d}\mathbf{r} \left( f(\psi) + \frac{1}{2}\kappa |\nabla \psi|^2 \right),
\end{equation}
with $\mathbf{r}$ the dimensionless spatial coordinate, $\kappa$ a constant often denoted as the `square-gradient coefficient', `gradient stiffness' or `elastic constant'.
Mass transport is driven by spatial gradients in the chemical potential as $\mathbf{j}=-M\nabla \mu$ \cite{deGrootBook84}, where  the chemical potential, $\mu \equiv \delta \mathcal{F}/\delta \psi$, is a functional derivative obtained from the Euler-Lagrange equation  $\mu = \partial f/\partial \psi -\kappa \nabla\cdot \partial |\nabla \psi|^2/\partial (\nabla \psi) = \partial f/\partial \psi - \kappa \nabla^2 \psi $.
Further, $M$ is the mobility, and is related to the cooperative diffusion coefficient as $D=M\partial \mu/\partial\psi$.

Taken into account that the order parameter is locally conserved  \footnote{For non-conserved order parameters, such as the orientation vector of liquid crystals (LC),  $\nabla\cdot M \nabla$ is replaced by $M$. In case of LCs, $M$ now represents a rotational rather than a translational mobility.}, its time evolution is described by $\partial \psi/\partial t = -\nabla \cdot \mathbf{j}$. By also adding a source term, $\alpha$, to describe e.g. a concentration quench \cite{Schaefer15, Schaefer16},  and a stochastic term, $\xi$ , that represents thermal fluctuations that obey the fluctuation-dissipation theorem, i.e., $\langle \xi(\mathbf{r},t)\rangle=0$ and $\langle \xi(\mathbf{r},t)\xi(\mathbf{r}',t')\rangle=-2\nabla\cdot M\nabla \delta(\mathbf{r}-\mathbf{r}')\delta(t-t')$ \cite{Cook70, Langer71},
we finally  have the modified Cahn-Hilliard-Cook (CHC) equation \cite{Cahn58, Cook70, Langer71, Hohenberg77,deGrootBook84} 
\begin{equation}
  \frac{\partial \psi}{\partial t} = \nabla \cdot \left(M \nabla\left[f_{\psi} - \kappa \nabla^2 \psi\right]\right) + \xi + \alpha.
\end{equation} 

In order to describe the time evolution of an emerging structure following a gradual quench into the miscibility gap of the phase diagram, we consider a system at time $t=0$ at the spinodal branch, that is, $\partial^2 f/\partial \psi^2 = 0$.
At this stage, the spatial variations of the order parameter are presumably small, i.e., $\psi=\psi^0 + \delta\psi$ with $|\delta\psi| \ll 1$.
Hence, we may write $\nabla f_\psi(t) \approx f_{\psi\psi}^0\nabla\delta\psi(t)$. 
Assuming a uniform (or absent) source term, Fourier transformation of the CHC equation gives 
\begin{equation}
  \frac{\partial \hat{\delta\psi}}{\partial t}(q,t) = -q^2 M \left(f_{\psi\psi}(t) + \kappa q^2 \right)\hat{\delta\psi}(q,t) + \hat{\xi}(q,t).
\end{equation}

For early times after the spinodal is crossed the second order derivative of the free energy is $f_{\psi\psi}(t) \approx -\Gamma t$, with 
\begin{equation}
  \Gamma \equiv \left.\frac{\partial f_{\psi\psi}}{\partial t}\right|_{t=t_\mathrm{spinodal}}
\end{equation}
 the quench rate, i.e., the rate by which the mixture is thermodynamically destabilised. Hence, the dominant structural wavenumber increases with time as $q_\ast(t) \approx \sqrt{(\Gamma /2\kappa)t}$, and the CHC equation reduces to 
\begin{equation}
\frac{\partial \hat{\delta\psi}_\ast}{\partial t/\tau} = \frac{M\Gamma^2}{4\kappa} \tau^3\left(\frac{t}{\tau}\right)^{2}\hat{\delta\psi}_\ast + \hat{\xi}\tau,
\end{equation}
with $\tau$ the characteristic ``time lag'' after which the phase-separated structure rapidly emerges ($\ln \delta\hat{\psi} \propto t^3$). 
This time scale, and the dominant wavenumber of the emerging structure, depend on the quench rate as \cite{Huston66}
\begin{equation}
  \tau \propto \Gamma^{-2/3}, \;\; q_\ast(\tau) \propto \Gamma^{1/6},
\end{equation}

\subsection{Microscopic lattice modelling: kinetic Monte Carlo}

The above free-energy functional is accurate if the binary mixture is sufficiently far from the critical point.
During a gradual quench, however, phase separation may take place close to the critical point and a microscopic model is needed.
The simplest approach is to model a binary $\mathrm{A}-\mathrm{B}$ mixture on a periodic $N \times N$   square lattice \cite{Kawasaki65, Carmesin86, Puri94, Glotzer95, Puri98, Cirillo19}. 
Each site is either vacant or occupied by $\mathrm{A}$ or $\mathrm{B}$.
The species have nearest-neighbour interactions $\epsilon_\mathrm{AA}$, $\epsilon_\mathrm{BB}$ and $\epsilon_\mathrm{AB}$.
Hence, for $N_\mathrm{AA}$, $N_\mathrm{BB}$ and $N_\mathrm{AB}$ nearest-neighbour pairs, the Hamiltonian is given by
\begin{equation}
  \mathcal{H} = \epsilon_\mathrm{AA}  N_\mathrm{AA}
+\epsilon_\mathrm{AB}  N_\mathrm{AB}
+\epsilon_\mathrm{BB}  N_\mathrm{BB}.
\end{equation}
At full surface coverage, i.e., for $\theta\equiv (N_\mathrm{A}+N_\mathrm{B})/N^2 = 1$, this Hamiltonian can be mapped onto the Ising Hamiltonian.
In this case, the components are driven to phase separate if $ \epsilon_\mathrm{AA}/\kB T, \epsilon_\mathrm{BB}/\kB T > \ln(1+\sqrt{2})/2 \approx 0.4407$ \cite{HuangBook87}. 
For lower concentrations, the critical temperature decreases.

A realistic example for which this lattice model would apply is the adsorption of molecules at the specific lattice sites of a metal surface \cite{HermseBook06, JansenBook12}.
In this case, surface diffusion of adsorbants is an activated process where a potential energy barrier has to be overcome. The hop rate then obeys an Arrhenius-type equation 
\begin{equation}
  W_\mathrm{hop} = \nu(T) \exp( -E_\mathrm{act}/\kB T ), \label{eq:Arrhenius}
\end{equation}
with $\nu$ the pre-exponential factor.
The temperature dependence of $W_\mathrm{hop}$ is predominated by the exponential term, and $\nu(T)$ is usually approximated to be isothermal \cite{JansenBook12}. 
The activation energy may be approximated by $\Ea=\Ea^0 + (1/2) \Delta\mathcal{H}$, with $\Ea^0$ the activation energy in the limit of low surface coverages and $1/2$ the `BEP coefficient' for simple hopping events \cite{HermseBook06}.

We model the response of the lattice fluid to a temperature quench as $T/T_0=1 - \Gamma_{T} t$, with $\Gamma_{T}$ a constant cooling rate and $T_0$ a temperature above the critical temperature.
Besides temperature quenches, we also consider isothermal concentration quenches.
For that purpose, we initialise the system using an empty surface and let molecular adsorption take place using the rates
\begin{equation}
  W_\mathrm{\theta, A\, or\, B} = \Gamma_\mathrm{\theta, A\, or\, B}.
\end{equation}
The resulting macroscopic adsorption rate is given by $\mathrm{d}\theta/\mathrm{d}t=(1-\theta)\Gamma_\mathrm{\theta}$, and the composition of the binary mixture is determined by the ratio between $\Gamma_\mathrm{\theta, A}$ and $\Gamma_\mathrm{\theta, B}$

In the present work, we simulate the dynamics using a variable-size-step method kinetic Monte Carlo (kMC) algorithm with a hierarchical selection method.~\cite{Lukkien98, JansenBook12}. 
This method is computationally efficient when the number of possible events is limited.  
For this reason, we use symmetric properties for species $\mathrm{A}$ and $\mathrm{B}$: We set the activation barriers equal and set the nearest-neighbour self-interactions to zero $\epsilon_\mathrm{AA}=\epsilon_\mathrm{BB}=0$. 

We analyse the results by calculating $\psi(k)$, with $k$ the lattice coordinate, and where $\psi(k)$ is $-1$, $1$ or $0$ if site $k$ is occupied by $\mathrm{A}$, $\mathrm{B}$, or neither.
We then obtain a structural length scale $R^*(t)$ by calculating the first minimum of the space-correlation function, and probe this quantity as a function of time  \footnote{
We calculate the typical length scale of a morphology at a given point in time as follows: First, we transform the order-parameter field, $\psi(\mathbf{r})$ to $q$ space using a $2$D Fourier transform, giving $\hat{\psi}(\mathbf{q})$. We then take the spherical/cylindrical average $S(q) = \langle |\hat{\psi}(\mathbf{q})|^2 \rangle$, and take the inverse Fourier transform to obtain the space-correlation function $C(r)$. The characteristic length scale that we measure is defined by the smallest distance $R_\ast$ where $C(r)$ takes its minimum value. For more detail, see Refs.~\cite{Singh12, Schaefer18}
}.

\subsection{Microscopic off-lattice modelling: molecular dynamics}

Complementary to the kMC simulations on a lattice, we perform two-dimensional Langevin dynamics simulations of a binary Lennard-Jones liquid with LAMMPS \cite{plimpton-1995} and its Kokkos package on a single NVidia GTX 1080 Ti.
The Lennard-Jones interaction potential between two particles $i$ and $j$ a distance $r_{ij}$ apart is given by
\begin{align*}
  U_{\mathrm{AB}}(r_{ij}) = 4 \epsilon_{\mathrm{AB}} \left[ (\sigma/r_{ij})^{12} - (\sigma/r_{ij})^6 \right]\times u(r_{\mathrm{c,AB}} - r_{ij}),
\end{align*}
with $u$ a unit step function: $u(x) = 1$ for $x>0$ and $0$ otherwise.
In the following, we express distances and energies in the Lennard-Jones units  $\sigma$ and $\epsilon$, respectively, and we express time in terms of the Langevin damping time $\tau$, i.e., the time it takes for the velocity auto-correlation to decay from $1$ to $\exp(-1)$.
We have set the parameter values to $\epsilon_{\mathrm{AB}} = \epsilon$ independent of the particle types, but $r_{\mathrm{c},11} = r_{\mathrm{c},22} = 3\sigma$ while $r_{\mathrm{c},12} = 2^{1/6}\sigma.$
Hence, the interactions between unlike particles is purely repulsive, while the interaction between like particles is attractive.
At our reduced Lennard-Jones density of $0.8$ particles per $\sigma^2$, this interaction potential drives  the phase separation at sufficiently low temperatures. 

We numerically solve the dynamics using a temporal step size of $0.005\tau$, and a planar simulation box with an area $A$ of both $(400\sigma)^2$ and $(800\sigma)^2$ to get a feel for finite size effects.
These simulations involve $128000$ particles for $A=(400\sigma)^2$ and $512000$ particles for $A = (800\sigma)^2$.
As an initial condition, we create a random packing of a single type of particles in the area, and then randomly select half of them and change them to the other type of particles.
We run each simulation for five random seeds.

To analyse the results using the same method as for the kMC simulations, we divide the simulation domain into $256\times256$ (for $A = (400 \sigma)^2$) or $512\times512$ (for $A=(800 \sigma)^2$) squares. We count for each of these squares $k$ the number of particles of type $\mathrm{A}$ and $\mathrm{B}$, which we denote as $N_\mathrm{A}(k)$ and $N_\mathrm{B}(k).$
Then, we assign to each square the order parameter $\psi(k) = 1$ if $N_\mathrm{A}(k) > N_\mathrm{B}(k),$ $\psi(k) = -1$ if $N_\mathrm{A}(k) > N_\mathrm{B}(k),$ or $0$ if $N_\mathrm{A}(k) = N_\mathrm{B}(k)$.
As before, the structural length scale then follows from the space-correlation function as before.

\section{Results}
 
In the following, we present the time evolution of the phase-separated morphologies of binary mixtures in response to a gradual composition quench using kMC simulations, and in response to a gradual temperature quench using both kMC and MD simulations. 
In section \ref{sec:universality} we combine the results and discuss their universality.


\begin{figure*}
  \centering
  \includegraphics*[width= \linewidth]{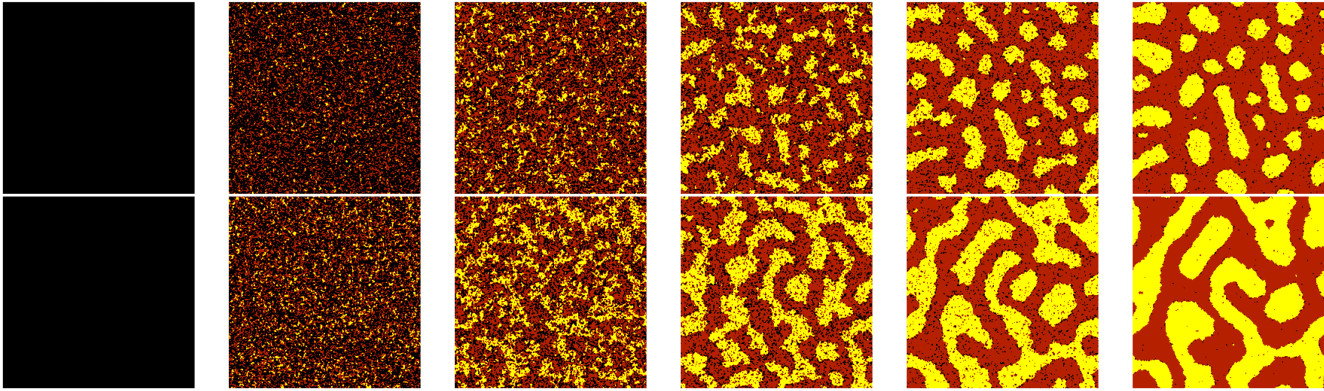}
  \caption{Time evolution of the structure during a concentration quench (left to right) for an off-critical (top) and a critical composition (bottom). The quench rate is $\Gamma_{\theta}/\nu = 3\cdot 10^{-7}$.}
  \label{fig:morph_theta_quench_time}
\end{figure*}


\subsection{Composition quenches (kMC)}

We have performed simulations of concentration quenches at both critical ($\Gamma_{\theta, \mathrm{A}}=\Gamma_{\theta, \mathrm{B}}$) and off-critical ($2\Gamma_{\theta, \mathrm{A}}=\Gamma_{\theta, \mathrm{B}}$) compositions, with rates varying from $\Gamma_{\theta} = \Gamma_{\theta, \mathrm{A}}+\Gamma_{\theta, \mathrm{B}} = 3\cdot 10^{-7}$ to $3\cdot 10^{-3}$ in units of the hop rate under dilute conditions, $\nu$.
In these simulations, the repulsive interaction energy between the two species was $\epsilon_{\mathrm{AB}}=3\kB T$.
Figure~\ref{fig:morph_theta_quench_time} shows representative morphologies of both a critical and an off-critical composition for the slowest quench.
In both cases, no structure seems discernible at early times.
As time proceeds, the surface coverage increases and the constituents phase separate via surface diffusion.
For the off-critical mixture, the morphology contains droplets and for the critical mixture it shows a bicontinuous-like structure.
Other than that, the time development of their characteristic length scales are similar.
At the late stages, the coverage reaches unity and the number of vacant sites for molecules to hop into decreases so that a final morphology `freezes' in.
This final structure becomes finer with an increasing quench rate.
This observation was seen experimentally as well, and was interpreted as a result of less time available for Ostwald ripening \cite{Franeker15}. 

\begin{figure}[!h]
  \centering
  \includegraphics*[trim={0cm 0 0 0}, width=8.2cm]{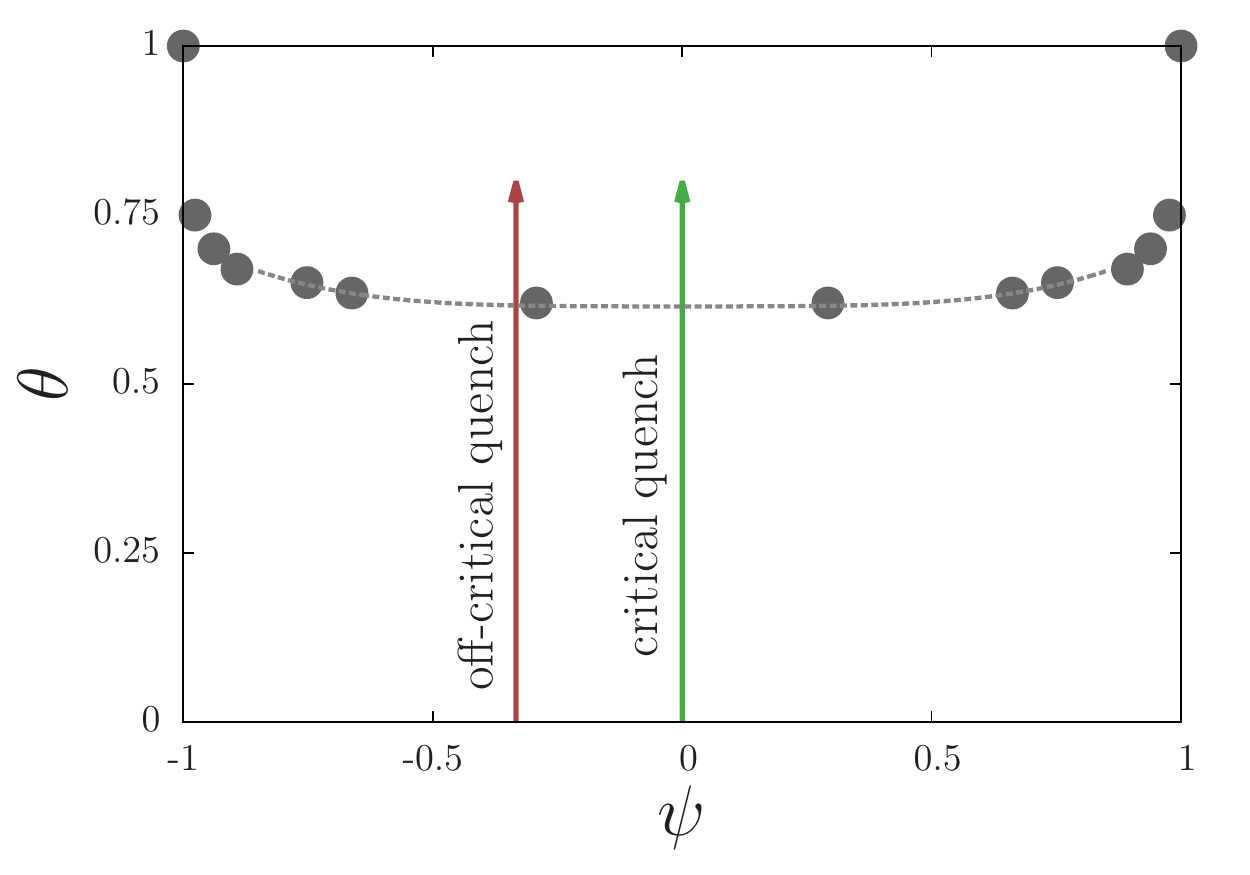}\hfill
  \includegraphics*[width=7.7cm]{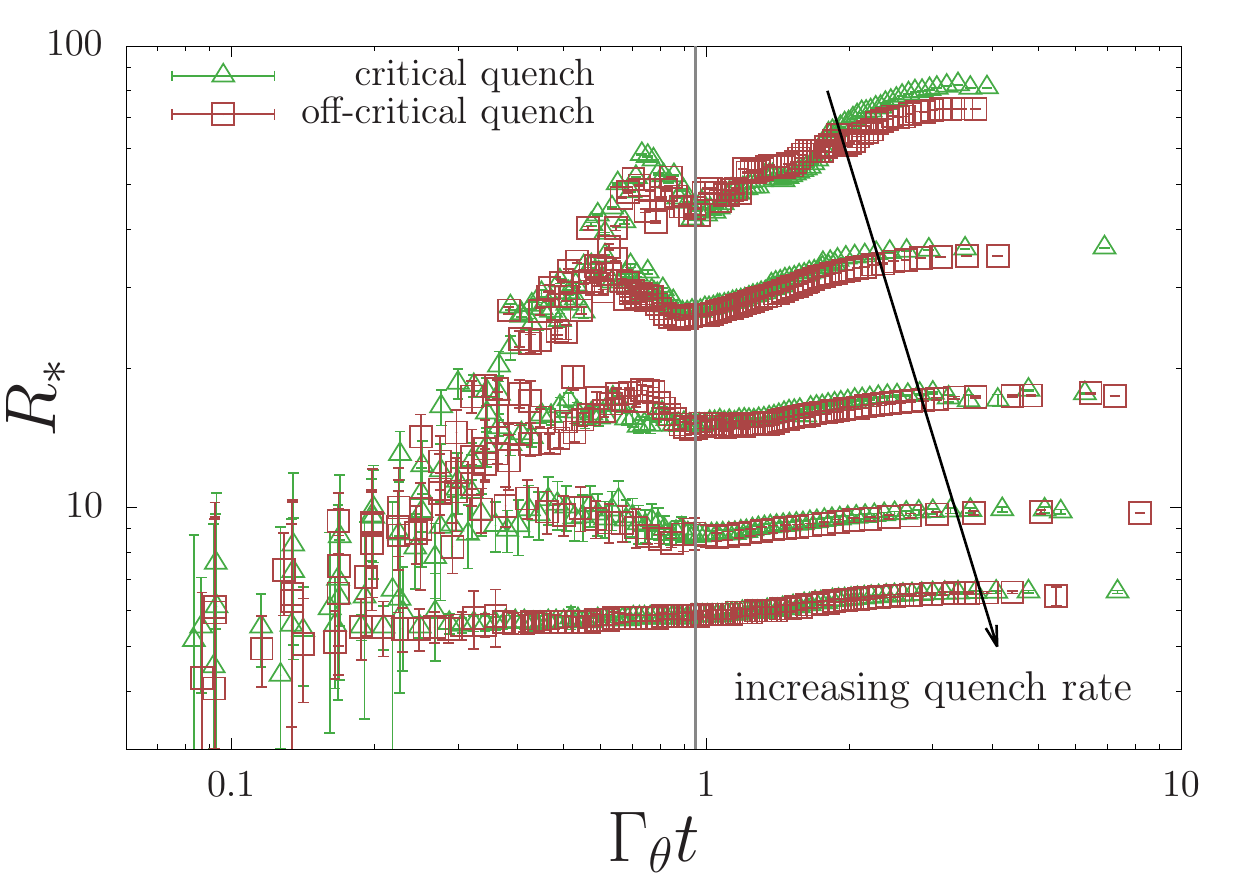}
  \caption{Top: Phase diagram for lateral interactions $\epsilon_\mathrm{AB}/\kB T=3$, with $\theta$ the surface coverage and $\psi$ the order parameter. The symbols are the binodal values as obtained by regular Monte Carlo simulations \cite{Schaefer18}. The dashed line is drawn as a guide to the eye. 
Bottom: Structural length scale $R_\ast$ as a function of scaled time $\nu_\mathrm{\theta} t$ for various quench rates $\Gamma\propto \nu_\mathrm{\theta}$. The vertical line indicates the time at which the binodal is crossed.
}
\label{fig:kmc_quench_result}
\end{figure}

Here, however, the time for structural coarsening is limited and the quench-rate dependence of the structural length is strongly affected by the early-stage structure development.
Crucially, with early-stage structure development we here refer to pre-transitional structuring prior to reaching the binodal, rather than spinodal decomposition after crossing the spinodal.
We address this effect by (i) determining the binodal concentration and (ii) by probing both the concentration and the characteristic length scale as a function of time.
We have determined the binodal concentration ($\theta\approx 0.63$) from the phase diagram in the top panel of Figure~\ref{fig:kmc_quench_result}  \footnote{We have calculated the phase diagrams using two isolated lattices with a fixed concentration, and an initialised composition. Using a standard Monte Carlo procedure, species are swapped between the two lattices with the appropriate acceptance probability. After sufficient steps, the compositions of the two lattices converge to the coexistence values \cite{Schaefer18}.}.
In that figure, the arrows indicate the trajectory of the examples of Figure~\ref{fig:morph_theta_quench_time}.

The structural length, $R_\ast$, as a function of time at those trajectories we present in the bottom panel of Figure~\ref{fig:kmc_quench_result} (the statistics are enhanced by performing simulations for $20$ random seeds per quench). 
The length scale is given in units of the lattice spacing, and time is rendered dimensionless using the quench rate. The vertical line indicates the time at which the unstable region of the phase diagram entered.
This figure shows that the morphology is  

\begin{figure*}
  \centering
  \includegraphics*[width= \linewidth]{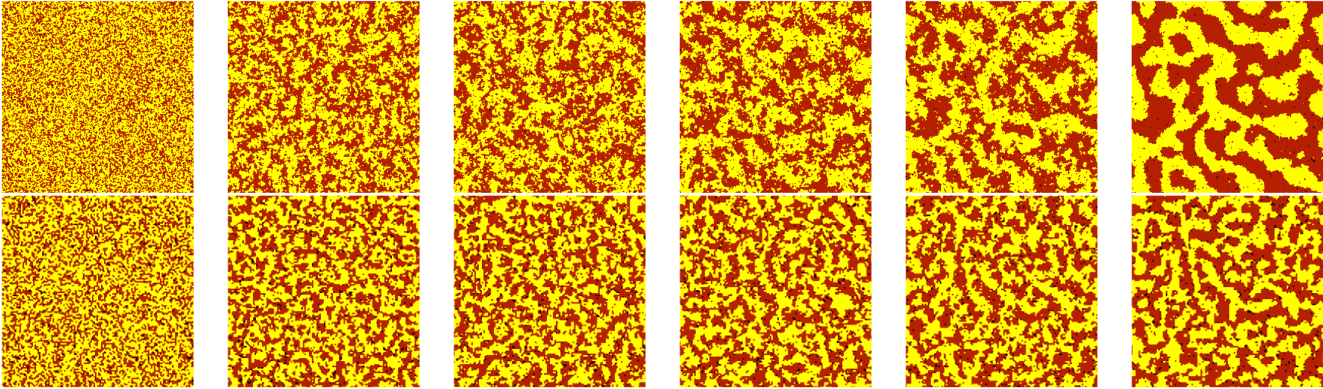}
  \caption{Time evolution (left to right) of the morphology during a slow (top, $\Gamma_{T} = 10^{-8}$) and a fast (bottom, $\Gamma_{T} = 10^{-6}$) temperature quench, as obtained from kMC simulations.  }
\label{fig:morph_T_quench_time}
\end{figure*}
\noindent finer for faster quenches not only when the system is frozen, but is finer already at the moment that the miscibility gap of the phase diagram is entered.
This implies that the pre-transitional structure development (i.e., in the single-phase region of the phase diagram) crucially affects the emerging morphology after crossing the binodal.  
The dependence of the structural length of the emerging structure on the quench rate is shown in Figure~\ref{fig:scaling_summary}, and will be discussed in section \ref{sec:universality}.

\subsection{Temperature quenches (kMC)}

Quantifying the early-stage structure development in composition-quench simulations is  somewhat obstructed by the noise imposed by the adsorption of species at random lattice sites.
This is prevented by keeping the concentration constant, and destabilising the mixture through cooling instead.
Indeed, in this section we perform kMC simulations of binary lattice mixtures that phase separate under gradual cooling.

We have chosen the parameter values in the simulations with applications in hetereogeneous metal catalysis in mind: For a lattice model to be physically realistic at the molecular scale, the activation energy for a hop of an adsorbant from one site to the other should always be larger than $\kB T$.
Assuming an activation energy that depends on the lateral interactions as mentioned below Eq.~\ref{eq:Arrhenius}, the interactions between species $\mathrm{A}$ and $\mathrm{B}$ should be sufficient weak.
Because of this Arrhenius-type activation barrier, the diffusivity decreases exponentially with a decreasing temperature.
Given these constraints, we have used surface coverages $\theta_\mathrm{A}=\theta_\mathrm{B}=0.475$, an activation energy $E_\mathrm{act}^0 = 4.2 \kB T_0$, and a repulsive interaction energy $\epsilon_\mathrm{AB}=0.8 \kB T_0$ at $t=0$. For these values, phase separation sets in when $\epsilon_\mathrm{AB}/\kB T > 1$, which is achieved by decreasing the temperature as $T/T_0=1 - \Gamma_{T} t$.
The cooling rate, $\Gamma_{T}$, ranges from $10^{-8}$ to $10^{-6}$ in units of the bare hop rate $\nu$.

\begin{figure}[h!]
  \centering
  \includegraphics*[width= 8 cm]{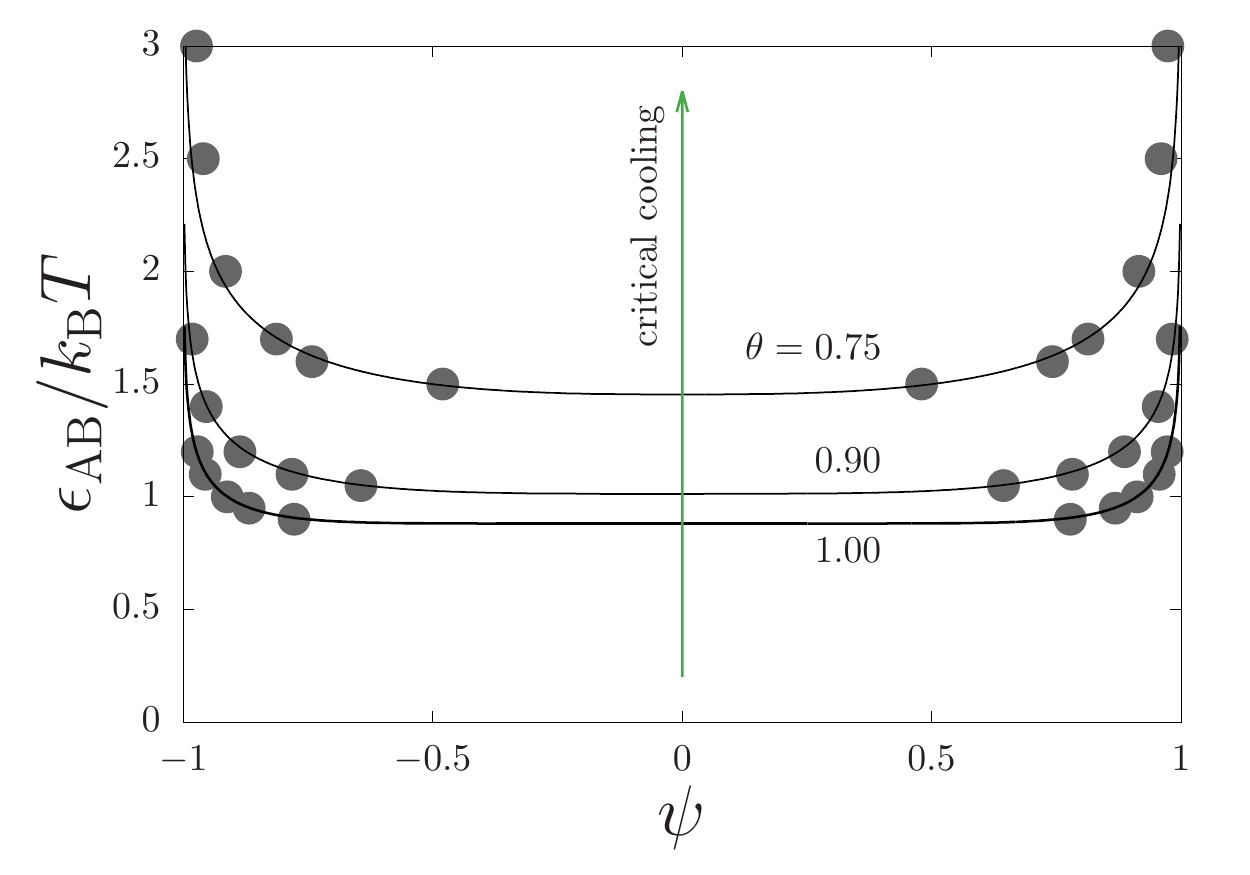}
  \includegraphics*[width= 8 cm]{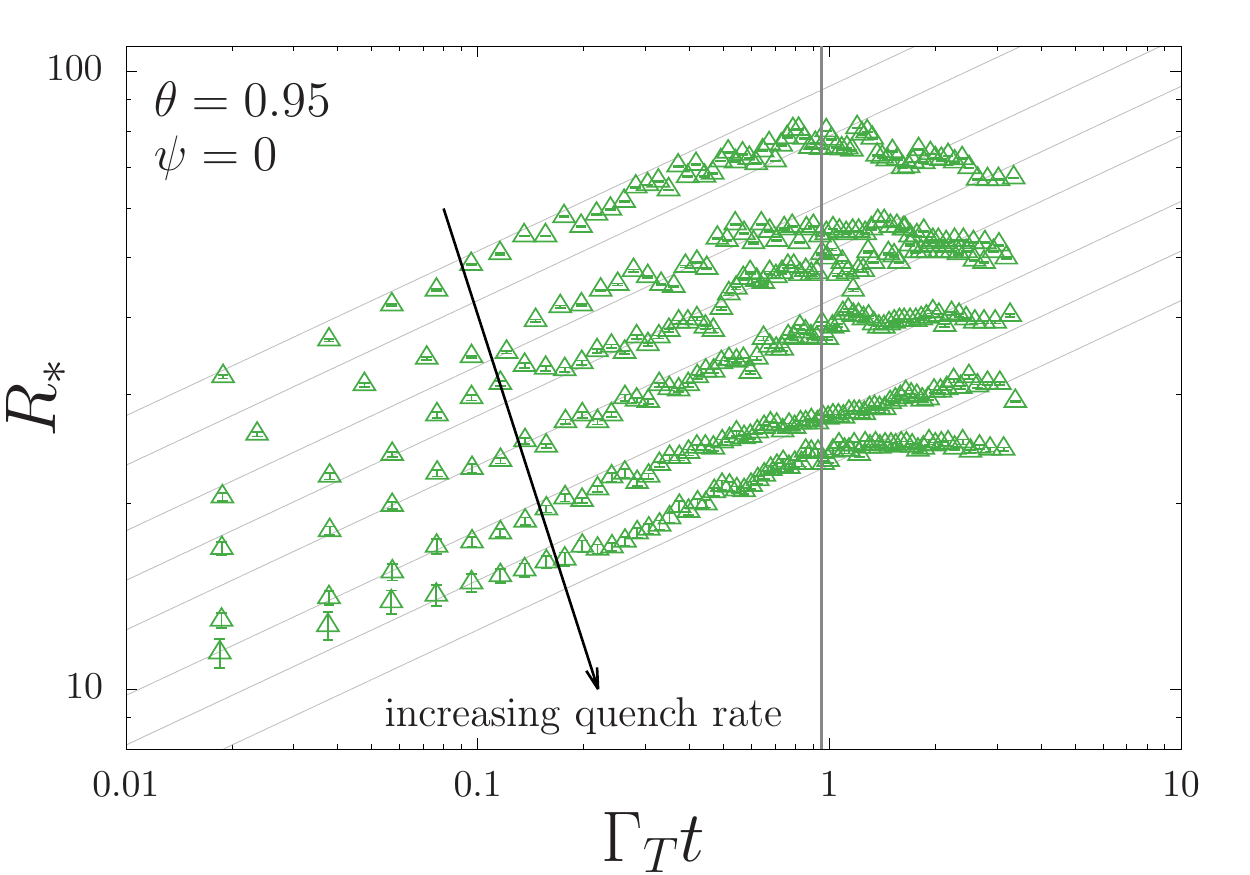}
  \caption{Top: Phase diagrams for various surface coverages $\theta$. $\epsilon_\mathrm{AB}$ is the nearest-neighbour interaction energy between species $\mathrm{A}$ and $\mathrm{B}$, $\kB T$ is the thermal energy, and $\psi$ is the order parameter. The symbols are the binodal values as obtained by regular Monte Carlo simulations \cite{Schaefer18}. The solid black line represents the Ising model, and the dashed lines are drawn as a guide to the eye. The arrows indicate temperature quenches. Bottom: Time evolution of the characteristic length scale, $R_\ast$, for various quench rates. The vertical line indicates the time at which the binodal is crossed.}
\label{fig:kmc_Tquench_result}
\end{figure}

In Figure~\ref{fig:morph_T_quench_time} we show representative examples of the morphology development during the fastest and the slowest quench.
In analogy with the concentration-quenched mixture, the final structure becomes coarser for slower quenches.
Like before, we distinguish between early- and late-stage structure development using the time at which the miscibility gap of the phase diagram is entered.
The top panel of Figure~\ref{fig:kmc_Tquench_result} shows the phase diagram for various surface coverages.
For the surface coverage of $\theta=0.95$ that we study here, the critical temperature is given by $\epsilon_\mathrm{AB}/\kB T\approx 1$, 
which for $\epsilon_\mathrm{AB}/\kB T_0=0.8$ is reached at time $t=0.2/\Gamma_{T}$.

\begin{figure*}
  \centering
  \includegraphics*[width= \linewidth]{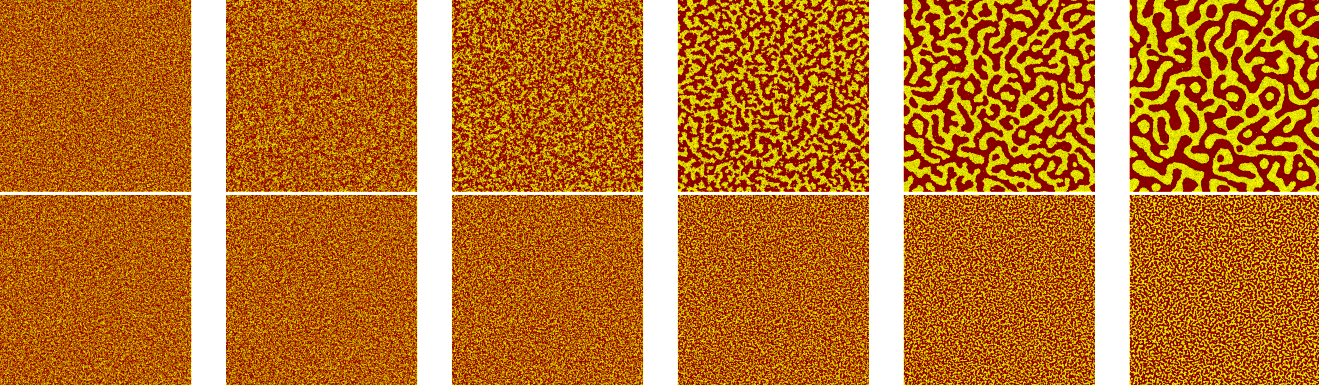}
        \caption{Time evolution (left to right) of the morphology during a slow ($\Gamma_T = 3.2~10^{-6}$, top) and a fast ($\Gamma_T = 3.2~10^{-4}$ bottom) temperature quench, as obtained from MD simulations. }
\label{fig:morph_MD}
\end{figure*}

This time is indicated by the vertical line in the bottom panel of Figure~\ref{fig:kmc_Tquench_result} .
This figure shows the time evolution of the characteristic length scale, $R_\ast$, of the morphologies (averaged over $20$ random seeds per quench).
If the temperature would be fixed at the critical temperature, one would expect the structural length scale to increase with the $4/15$ power of time \cite{Hohenberg77, Alexander94, Godreche04}.
Despite the fact that our simulations have a transient temperature, we do recover this power law (grey lines) for early times.
After entering the miscibility gap, the purity of the phase separated domains increases (see Figure~\ref{fig:morph_T_quench_time}), but the size of the domains prompty arrests as the mixture freezes at low temperatures.
The dependence of the structural length of the emerging structure on the quench rate is shown in Figure~\ref{fig:scaling_summary}, and will be discussed in section \ref{sec:universality}.

\subsection{Temperature quenches (MD)}

The constraints imposed by the strong temperature-dependence of the mobility as well as the geometry of the lattice are lifted
 in a natural way by resorting to (Langevin) molecular dynamics (MD) simulations.
Indeed, we have conducted MD simulations of a binary mixture with $\epsilon_\mathrm{AB} = 1$ in Lennard-Jones unit energy $\epsilon$.
In these simulations, we have initialised the structure at temperature   $T_0 = 2.5\epsilon$, and have decreased the temperature to $T_\mathrm{final}/T_0 = 1-\Gamma_{T} t_\mathrm{final} = 0.2$ linearly in time with $\Gamma_{T}$ varying from $3.2\cdot 10^{-4}$ to $3.2\cdot 10^{-6}$.

Figure~\ref{fig:morph_MD} shows representative images of the time evolution of the morphology for quench rates of  $3.2\cdot 10^{-4}$ and  $3.2\cdot 10^{-6}$.
These morphologies are qualitatively similar to those found in the kMC simulations of temperature quenched binary mixtures (see Figure~\ref{fig:morph_T_quench_time}).
Indeed, some early-stage coarsening is visible, as well as purification of the domains at late times.
In contrast to the kMC simulations, the constituents remain mobile and coarsening proceeds at the late stages.
By fitting the late-stage coarsening exponent, $R_\ast \propto (t-\tau)^{1/3}$, with $\tau$ the time at which late-stage coarsening sets in, we extract $\tau$ and estimate the critical temperature below which phase separation takes place.

\begin{figure}[h!]
  \centering
  \includegraphics*[width= 8 cm]{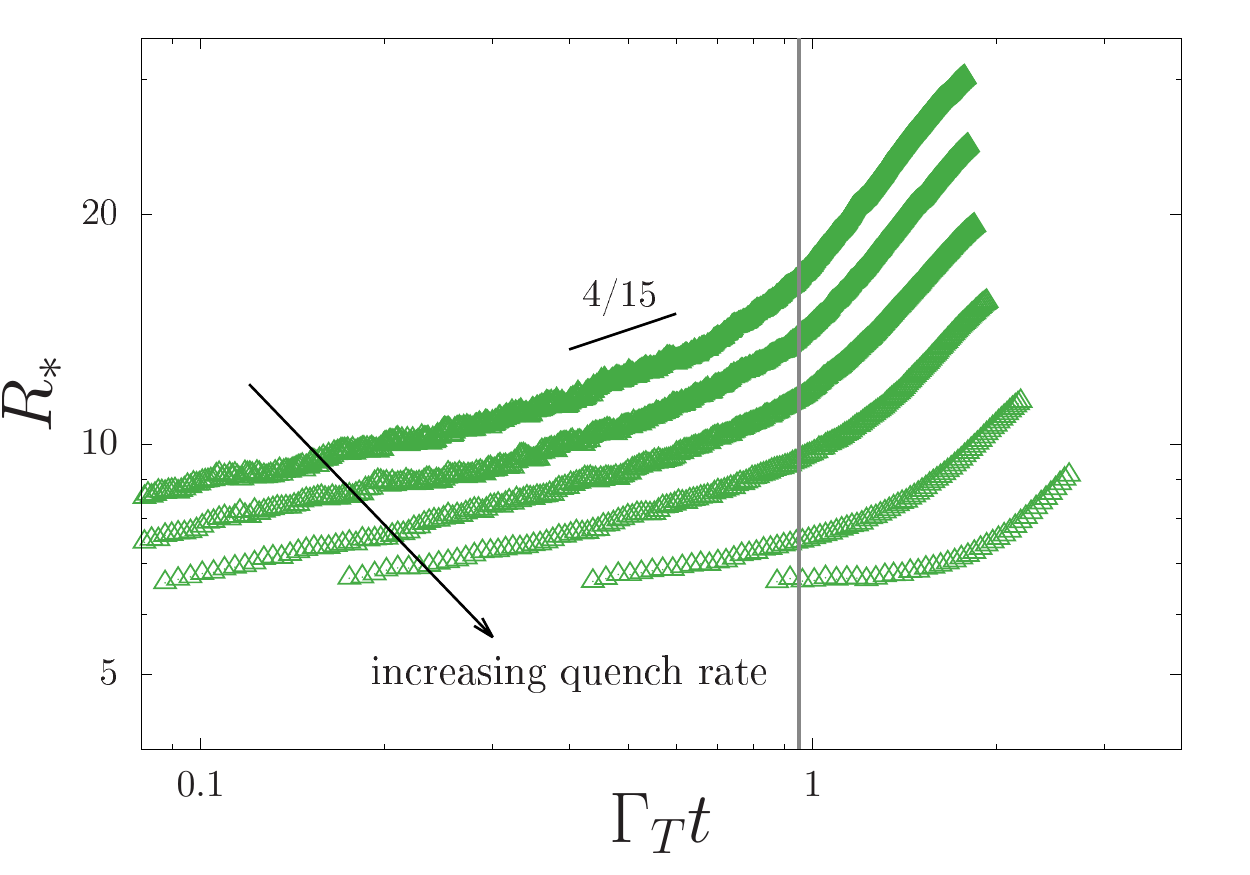}
  \caption{Time evolution of the characteristic length scale, $R_\ast$, for quench rates ranging from $3.2\cdot 10^{-6}$ to $3.2\cdot 10^{-4}$. The vertical line indicates the (approximate) time at which the binodal is crossed.}
\label{fig:MD_Tquench_result}
\end{figure}

The time at which this happens is indicated by the vertical line in Figure~\ref{fig:MD_Tquench_result}.
This figure shows the characteristic length scale, $R_\ast$, (in units of Lennard-Jones length scale $\sigma$) as a function of time. 
We have rendered the time dimensionless using a quench rate that we estimated by comparing the time scale of phase separation in Figure \ref{fig:MD_Tquench_result} to those in Figure~\ref{fig:kmc_Tquench_result}.
Like before, $R_\ast$ increases at early times reasonably consistently with the $4/15$th power of time up to the time where the critical temperature is reached.
After that, true phase separation and coarsening of the emerging structure sets in.

\subsection{Universal scaling}\label{sec:universality}

\begin{figure}[ht!]
  \centering
  \includegraphics*[width= 8 cm]{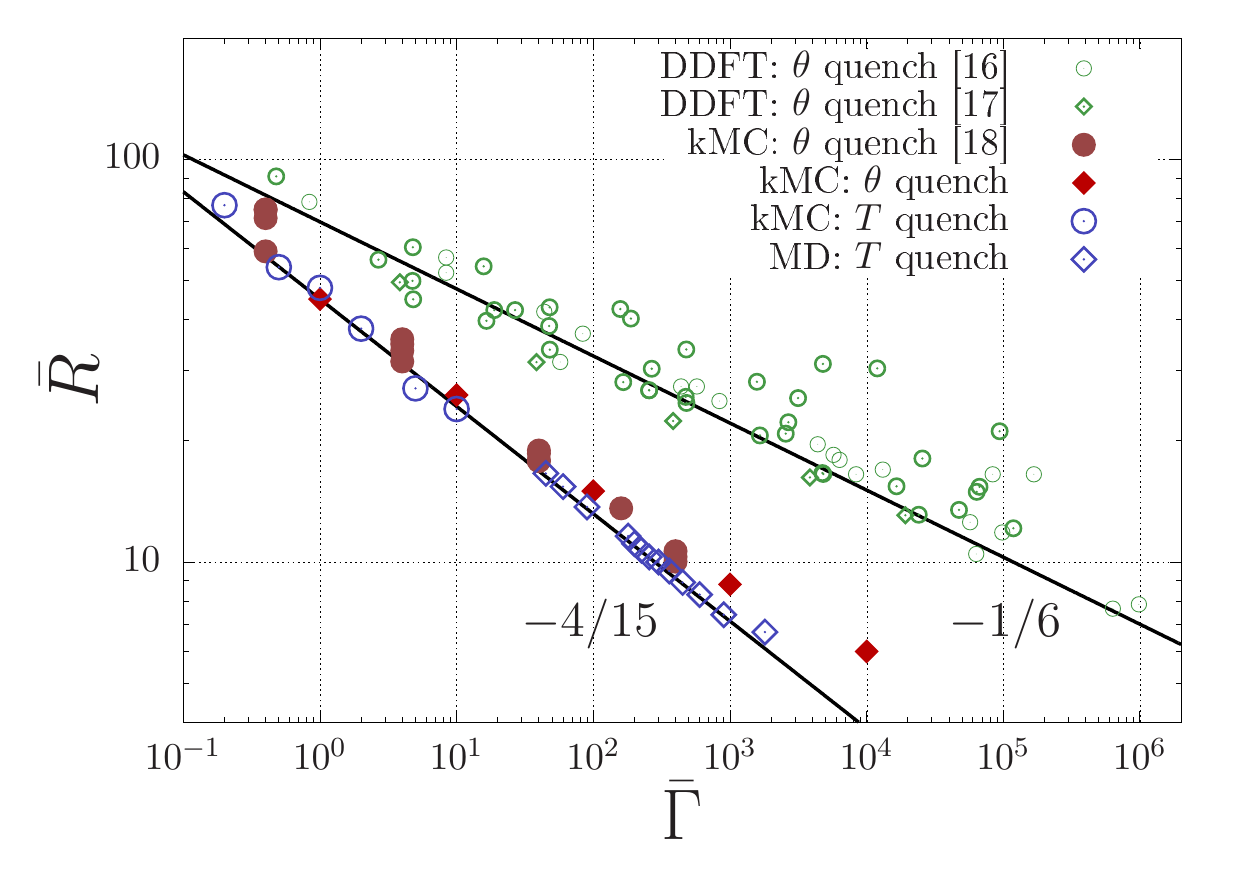}
  \caption{Dimensionless length scale, $\bar{R}$, of emerging structures as a function of the dimensionless quench rate, $\bar{\Gamma}$, as calculated using density functional theory (DDFT) \cite{Schaefer15, Schaefer16}, kinetic Monte Carlo (kMC) lattice simulations \cite{Schaefer18}, as well as Langeving molecular dynamics (MD) simulations.
The results where collected for temperature ($T$) and concentration ($\theta$) quenches and for various numbers of components; see main text for more detail.
The lines with slopes $-1/6$ and $-4/15$ represent the expected results from mean-field theory \cite{Huston66} and critical dynamics \cite{Hohenberg77}, respectively.  
  }
  \label{fig:scaling_summary}
\end{figure}

For all three types of simulations of binary mixtures above, the emerging structure is determined by the correlation length, $\bar{R}\equiv R_\ast(\tau)$, at the point in time, $\tau$, where the miscibility gap of the phase diagram is entered.
In Figure~\ref{fig:scaling_summary} we plot this quantity as a function of the (scaled) quench rate for all simulations, and compare this to the results of earlier reported kMC simulations of a single-component phase-separating mixture \cite{Schaefer18} and for dynamic density functional theory (DDFT) simulations of phase separating binary \cite{Schaefer15} and ternary \cite{Schaefer16} mixtures.
All of those reported works studied phase separation in response to a concentration rather than a temperature quench.
The $\bar{R}$ values of the DDFT results, as well as the quench rates $\bar{\Gamma}$ have been shifted for clarity.

From Figure~\ref{fig:scaling_summary} we see that the microscopic modelling results using kMC and MD show a much stronger dependence ($-4/15$) on the quench rate than the DDFT calculations do ($-1/6$).
While the power $-1/6$ is expected from mean-field theory (see section~\ref{sec:mean-field}),
we explain the power $-4/15$ using universal critical dynamics: 
Prior to reaching the critical concentration or temperature, the correlation length increases with the $4/15$th power of time, and the time available for coarsening is inversely proportional to the quench rate.

\section{Conclusions}

In this work, we have theoretically investigated the structuring of a binary fluid in response to a homogeneous concentration and temperature quench.
Using kinetic Monte Carlo (kMC) and molecular dynamics (MD) simulations,
we have discussed the failure of mean-field theory under conditions of a gradual quench.
Neither the thermodynamic driving force for phase separation, nor the collective diffusion of the constituents determines the emerging structure.
Instead, the emerging structure is determined solely by the mobility of the constituents at the pre-transitional stages of the quench.
During these stages, the structure `ages' via critical dynamics and the correlation length increases with the $4/15$th power of time \cite{Hohenberg77, Alexander94, Godreche04}.  
Because the quench rate sets the time available for pre-transitional structuring, the emerging morphology is characterised with a length scale that decreases with the $-4/15$th power of the quench rate, rather than the much weaker $-1/6$th power that is expected from mean-field theory.


While these findings are strictly limited to two-dimensional binary fluids with local conservation laws (i.e., `Passive Model B') \cite{Kawasaki65}, our conclusions can be extrapolated to various three-dimensional examples with different types of critical dynamics \cite{Hohenberg77}, see Table~\ref{tab:scaling}. 
Indeed, some  applications in soft-matter science with Model A dynamics include reaction-diffusion mixtures with (reversible) chemical reactions  \cite{Puri94, Puri98, Glotzer95, Furtado06, Singh12, Krishnan15}, as well as liquid crystals \cite{Dhont05, Lettinga06, Green09}.
Phase-separation phenomena in active matter (i.e., mixtures of active / self-propelling particles) are governed by `Active Model B' dynamics \cite{Kardar86, Wittkowski14}.  
From Table~\ref{tab:scaling}, we expect that in both Model A and Active Model B kinetics the quench rate is even of stronger influence to the emerging morphology than in the Passive Model B kinetics examples that we have studied so far. 
We hope that the present work will assist the experimental and theoretical research to phase-separation phenomena under transient conditions in the fields of both biophysics  and materials science.

\begin{table}[!ht]
\small
  \caption{\ Values of $1/z$ for power law $R_0 \propto \Gamma^{1/z}$,\textendash\ with $R_0$ the emerging \\ length scale of phase-separating morphologies and $\Gamma$  the quench rate. \\ We have studied Passive Model B in $2$D in the present work, the other \\ values are as expected from dynamic critical scaling arguments \cite{Hohenberg77}}
  \label{tab:scaling}
  \begin{tabular*}{0.48\textwidth}{@{\extracolsep{\fill}}ccc}
    \hline
     &   $2$D & $3$D\\
    \hline
   Model A &         $0.46$  & $0.49$  \\
   Passive Model B  &  $4/15\approx 0.2666$    & $0.253$ \\
   Active Model B & $0.62$  & $0.5$ \\
    \hline
  \end{tabular*}
\end{table}

\section*{Acknowledgements}
C.S. and T.C.B.M acknowledge the Engineering and Physical Sciences Research Council (Grant No. EP/N031431/1) for funding.
S.P. was supported by Award Number R01GM108021 from the National Institute Of General Medical Sciences and the Brandeis Center for Bioinspired Soft Materials, an NSF MRSEC,  DMR-1420382. Computational resources were provided by NSF XSEDE computing resources TG-MCB090163, and the Brandeis HPCC which is partially supported by DMR-1420382. S.P. also thanks Michael F. Hagan for valuable discussions.


\balance


\end{document}